\documentclass[aip,apl,reprint,twocolumn,showpacs]{revtex4-1}
\usepackage{graphicx}
\usepackage{graphicx, subfigure}
\usepackage{wasysym}
\usepackage{amssymb}
\usepackage{color}
\usepackage{natbib}
\usepackage{bm}
\usepackage{amsfonts}

\setcitestyle{super}
                        

\def\urlprefix{}
\def\url#1{}

\begin{document}

\title{Thermal conductivity of group-IV Semiconductors from a Kinetic-Collective Model}

\author{C. de Tomas}
\affiliation{Department of Physics, Universitat Aut\`onoma de Barcelona, 08193 Bellaterra, Catalonia, Spain}

\author{A. Cantarero}
\affiliation{Materials Science Institute, University of Valencia, P. O. Box 22085, 46071 Valencia, Spain}

\author{A. F. Lopeandia}
\affiliation{Department of Physics, Universitat Aut\`onoma de Barcelona, 08193 Bellaterra, Catalonia, Spain}

\author{F. X. Alvarez}
\email{xavier.alvarez@uab.cat}
\affiliation{Department of Physics, Universitat Aut\`onoma de Barcelona, 08193 Bellaterra, Catalonia, Spain}
\email{xavier.alvarez@uab.cat}

\date{\today}

\begin{abstract}
The thermal conductivity of several diamond-like materials is calculated from a kinetic-collective model. From this approach, a thermal conductivity expression is obtained that includes a transition from a kinetic (free) to a collective (hydrodynamic) behavior of the phonon field. The expression contains only three parameters. Once fitted to natural occurring silicon, the same parameters for the other materials are directly calculated from theoretical relations. The results are in good agreement with experimental data.
\end{abstract}
\maketitle

A general model able to predict the thermal conductivity at all ranges of temperature for different materials and device sizes has been extensively debated in the literature. For this, Boltzmann Transport Equation (BTE) is the usual starting point, but obtaining a general predictive solution for thermal transport is a tremendous task that at the moment has not been completely achieved. 

Most of the works proposed in the last decades agree that the cause of this difficulty is the presence of momentum conservative phonon-phonon scattering (N-processes).\cite{Fugallo2013,Allen2013,Callaway1959,Guyer1966,Broido2007,Asen-Palmer1997} This scattering mechanism moves the distribution function from equilibrium and when it is the dominant process, kinetic approaches based on the calculation of the conductivity per mode fail. In order to deal with this, two different strategies are used to obtain some approximate solutions of the BTE: numerical\cite{Esfarjani2011,Fugallo2013,Li2013} and phenomenological models.\cite{Morelli2002,Mingo2003,Chantrenne2005,Allen2013,Alvarez2010}

An example of the numerical approach is to combine ab-initio calculations for the relaxation terms with a recursive method of solution for the BTE. This has led to valuable results for semiconductor bulk samples in the temperature range $[50,400]$  K\cite{Broido2007,Ward2009}. The goal of these works is to obtain the largest range of predictions without using any fitting parameters. The main problem for this kind of approaches is the computational cost at low temperatures or in reduced size samples. 

On the other side, phenomenological approaches try to obtain simplified expressions for the thermal conductivity based on applying a minimum set of simplifications to the BTE. Their main goal is not to predict accurately  experimental values, but to obtain a reasonable approximation that allows us to interpret the results in terms of physical phenomena. One of the most well-known phenomenological models is the one proposed by Callaway \cite{Callaway1959}(CM), but in the last years some issues related to the prediction of reduced-size samples has led to explore some modifications\cite{Kazan2010,Mingo2003,Allen2013}. In a forthcoming paper, Tomas et al.\cite{DeTomas2013} propose an alternative kinetic-collective model (KCM) improving predictions for silicon (Si) thermal conductivity. The strong point is that they establish two regimes of phonon thermal transport depending on the dominance of normal or resistive scattering, this is the collective and the kinetic regime, respectively.

Although CM and KCM seem similar, the more detailed nature of the second proposal makes it a more suitable starting point to study phononic thermal transport in crystals. The model is not only a useful tool to predict thermal conductivity of different shape, size or isotopic composition samples but also gives insight into the physics underlying phonon-phonon interactions in thermal transport. The agreement with experimental data from bulk to nanoscale Si samples in the whole range of temperatures is remarkable.\cite{DeTomas2013} This good result with Si has motivated us to check the robustness of the model by applying it to other materials than Si. The aim of this work is to predict the thermal conductivity of a set of samples of several materials without introducing any extra fitting parameter. In order to focus only on the relaxation times and dismiss the effects coming from the crystal structure, we have studied only diamond-like crystals, specifically the family of group-IV semiconductors: diamond (C), Si, germanium (Ge) and grey tin ($\alpha$-Sn). Their dispersion relations have been calculated using the Bond charge model.\cite{Weber1974}

In general, thermal conductivity $\kappa$ can be obtained as a combination of a kinetic $\kappa_{\mathrm{kin}}$ and a collective $\kappa_{\mathrm{coll}}$ contribution \cite{DeTomas2013}
\begin{equation}
\label{kappa_tot}
  \kappa=\kappa_{\mathrm{kin}}(1-\Sigma)+ \kappa_{\mathrm{coll}}\Sigma
\end{equation}
where 
\begin{equation}
\label{conductivity_kin_dos}
\kappa_{\rm{kin}}=\frac{1}{3}\int \hbar \omega \tau v_g^{2} \frac{\partial f^0}{\partial T} Dd\omega
\end{equation}
and
\begin{equation}\label{conductivity_coll_dos}
\kappa_{\mathrm{coll}}=\frac{1}{3}\frac{\left(\int v_{g} q \frac{\partial f^0}{\partial T}Dd\omega\right)^{2}}{\int \frac{q^2}{\hbar \omega}\frac{1}{\tau} \frac{\partial f^0}{\partial T}Dd\omega}.
\end{equation}
with $\omega$ the phonon frequency, $q$ wave vector, $f^0$ equilibrium distribution, $D$ density of states, $v_g$ group velocity, and $\tau$ relaxation time. The weight of these contributions depends on a switching factor $\Sigma=\left[1+\langle\tau_N\rangle/\langle\tau_R\rangle\right]^{-1}$ that takes values in the range $\Sigma \in [0,1]$ depending on the averages of normal $\langle \tau_{N}\rangle$ and resistive $\langle \tau_{R}\rangle$ relaxation times, where
\begin{equation}\label{k_kin_int}
\langle \tau_{i}\rangle=\frac{\int \hbar \omega \tau_{i} \frac{\partial f^0}{\partial T}Dd\omega}{\int \hbar \omega \frac{\partial f^0}{\partial T}Dd\omega}.
\end{equation}
Expression (\ref{kappa_tot}) describes  a transition from a kinetic regime when resistive scattering is dominant, to a collective behavior when N-process is dominant. For the set of group-IV samples we propose to use the same expressions for the relaxation times as those used for silicon. For the boundary term we use\cite{Casimir1938} 
$\tau^{-1}_{B}=v_g/L_{\mathrm{eff}}$ where $L_{\mathrm{eff}}=1.12\sqrt{A}$ is the effective size for a sample with square cross-section $A$. For impurity scattering we use\cite{Klemens1955, Tamura1983} $\tau^{-1}_{I}=(V\Gamma/4\pi v_g^{3})\omega^4$, being $V$ the atomic volume and $\Gamma= \sum_i c_i\left(\Delta M/M\right)^2 $ the mass-fluctuation factor, with $c_i$ the isotopic fraction. Values of sizes and $\Gamma$ for the different samples appear in Table \ref{table:gamma}.
\begin{table}
\caption{Values of $\Gamma$, cross-section $A$ and effective size $L_{\mathrm{eff}}$ for natural occurring and different isotopic composition Si, Ge, C and $\alpha$-Sn samples.\cite{Inyushkin2004,Asen-Palmer1997,Wei1993,Onn1992}}
\begin{tabular}{ c c c c c }
\hline
\hline
Sample & $\Gamma$ & A (mm$^2$)& $L_{\mathrm{eff}}$(mm) & Reference\\
\hline
$^{\rm na}$Si & $20.01\times 10^{-5}$ & $2.00\times 3.12$ &2.8 & Ref.\cite{Inyushkin2004}\\ 
$^{28}$Si & $3.2\times 10^{-7} $ & $2.00\times 3.12$ &2.8 & Ref.\cite{Inyushkin2004}\\
$^{\rm na}$Ge & $58.7\times 10^{-5}$ &$2.46\times 2.50$ &2.78 & Ref.\cite{Asen-Palmer1997}\\ 
$^{70/76}$Ge &	$1.53\times 10^{-3}$ & $2.02\times 2.00$&2.25 & Ref.\cite{Asen-Palmer1997}\\
$^{70}$Ge99.99 &	1.8$\times 10^{-7}$ & $2.20\times 2.50$&2.63 & Ref.\cite{Asen-Palmer1997}\\
$^{70}$Ge96.3 &	$7.57\times 10^{-5}$ & $2.50\times 2.50$&2.8 & Ref.\cite{Asen-Palmer1997}\\
$^{\rm na}$C & $7.54\times 10^{-5}$ & $\AC 1\times 1$& 0.82 & Ref.\cite{Onn1992}\\  
$^{13}$C & $6.94\times 10^{-6}$ & $1\times 2$&1.58 &  Ref.\cite{Wei1993} \\
$^{\rm na}$Sn & $33.46\times 10^{-5}$ &  $2.5\times 2.5$& 2.8 & -\\
\hline
\end{tabular}\label{table:gamma}
\end{table}

For N-processes we use
\begin{equation}
\tau_{N}=\frac{1}{B _{N}T}+\frac{1}{B'_NT^{3}\omega^2[1-\exp(-3T/\Theta_D)]}\quad,
\label{tauN}
\end{equation}
and for umklapp (U-processes)
\begin{equation}
\tau_U=\frac{\exp(\Theta_U/T)}{B_U\omega^4T[1-\exp(-3T/\Theta_D)]},
\label{tauU}
\end{equation}
where $\Theta_D$ is the Debye temperature and $\Theta_U$ is the umklapp extinction temperature calculated from the dispersion relations\cite{DeTomas2013}. Expressions (\ref{tauN})-(\ref{tauU}) agree to those provided by Ward and Broido obtained by ab-initio calculations in the temperature range where they compare to experimental data [50-300]K.\cite{Ward2010}

Although (\ref{tauN})-(\ref{tauU}) include some parameters, according to Leibfried and Schl\"omann\cite{Leibfried1954}, there is a semiempiric expression to obtain $B_U$ and $B_N$. Later recovered by Morelli et al.\cite{Morelli2002}, this expression is written for both parameters
\begin{equation}
B_{U/N} \approx \left(\frac{K_B}{\hbar}\right)^b \frac{\gamma^2\hbar V^{(a+b-2)/3}}{M v_g^{a+b}}
\label{BN}
\end{equation}
where $K_B$ is the Boltzmann constant, $\hbar$ the Planck constant, $\gamma$ the Gr\"uneisen parameter, $M$ the atomic mass, being $a,b$ the exponents of frequency and temperature dependence respectively, since $\tau^{-1}_{U/N}\propto\omega^aT^b$.

From our expressions it can be seen that $a+b=5$ for $B_{U/N}$ and $a+b=1$ for $B'_{N}$. 
Excluding $K_B$ and $\hbar$, we find that the value of the parameters is exclusively related to four magnitudes: $\gamma, v_g, V$ and $M$, which are characteristic of each material (see Table \ref{table:properties}). The same mean value $\gamma=0.7$ can be used for all the group-IV materials as indicated by Slack \cite{Slack1973}. For $v_g$ we calculate the value at the zone-center of the Brillouin zone from dispersion relations using $v=\left[2/3v_T+1/3v_L\right]^{-1}$.

Expression (\ref{BN}) allows to express the $B$ parameters of all the materials in terms of the values of one of them. 
We take Si as the reference material, the values $B_{U,Si}=3.0\times 10^{-46}$ s$^3$K$^{-1}$, $B_{N,Si}=2.6\times 10^{-23}$ sK$^{-3}$ and $B'_{N,Si}=4.8\times 10^8$ s$^{-1}$K$^{-1}$ provide the best fit for naturally occurring Si ($^{\rm na}$Si). Now we calculate the respective values for the other of the materials as $B_{U/N,x}=f_{x} B_{U/N,Si}$ and $B'_{N,x}=f'_{x} B'_{N,Si}$ where $x$ denotes the material, and 

\begin{equation}
\label{BUlow}
f_{x}= \frac{\left[\gamma^2 V/M v^5\right]_{x}}{\left[\gamma^2V/M v^5\right]_{Si}}
\,\,
; \,\,\,f'_{x}= \frac{\left[\gamma^2/M v V^{1/3}\right]_{x}}{\left[\gamma^2/M v V^{1/3}\right]_{Si}}
\end{equation}
are the conversion factors for the low  and high temperature parameters, $B_{U/N}$ and  $B'_N$ respectively. Calculated values of $f_{x}$ and $f'_{x}$ are in Table \ref{table:factor}.

\begin{table}[ht]
\caption{Materials properties: Debye temperature $\Theta_D$, atomic mass $M$, atomic volume $V$, mean zone-center velocity $v_g$ and umklapp extinction temperature $\Theta_U$. } 
\centering 
\begin{tabular}{c c c c c c}
\hline 
\hline
 Material&$\Theta_D$ (K) & $M$ (g mol$^{-1}$) &$V$ (m${^3}$) & $v_g$ (m/s)& $\Theta_U$ (K) \\[0.5ex]
\hline
Si & 645 			&28.086 	&20.01$\times 10^{-30}$ 	& 6272 	&126 \\
Ge &  375 			& 72.63 	& 22.75$\times 10^{-30}$ 	& 3923	&70\\
C &  1850 			& 12 		& 5.67$\times 10^{-30}$ 	&13746 	& 405\\
$\alpha$-Sn &230 	& 118.69	&34.05$\times 10^{-30}$ 	& 2769	&37 \\
\hline
\end{tabular}\label{table:properties}
\end{table}

\begin{table}[ht]
\caption{Conversion factors $f_x$ and $f'_x$ calculated from Eqs. (\ref{BUlow}) for each material.}
\centering
\begin{tabular}{c c c c c c }
\hline
\hline
Parameter & Factor & $\alpha$-Sn & Ge & Si & C \\ [0.5ex]
\hline
$B_{U/N}$ & $f_{x}$ & 24.42 & 5.60 & 1 & 0.023 \\ 
$B'_{N}$ & $f'_{x}$ & 0.45 & 0.618 & 1 & 1.091 \\ 
\hline
\end{tabular}\label{table:factor}
\end{table}

\begin{figure*}[ht!]
     \begin{center}
            \label{fig:FIT}
            \includegraphics[width=\textwidth]{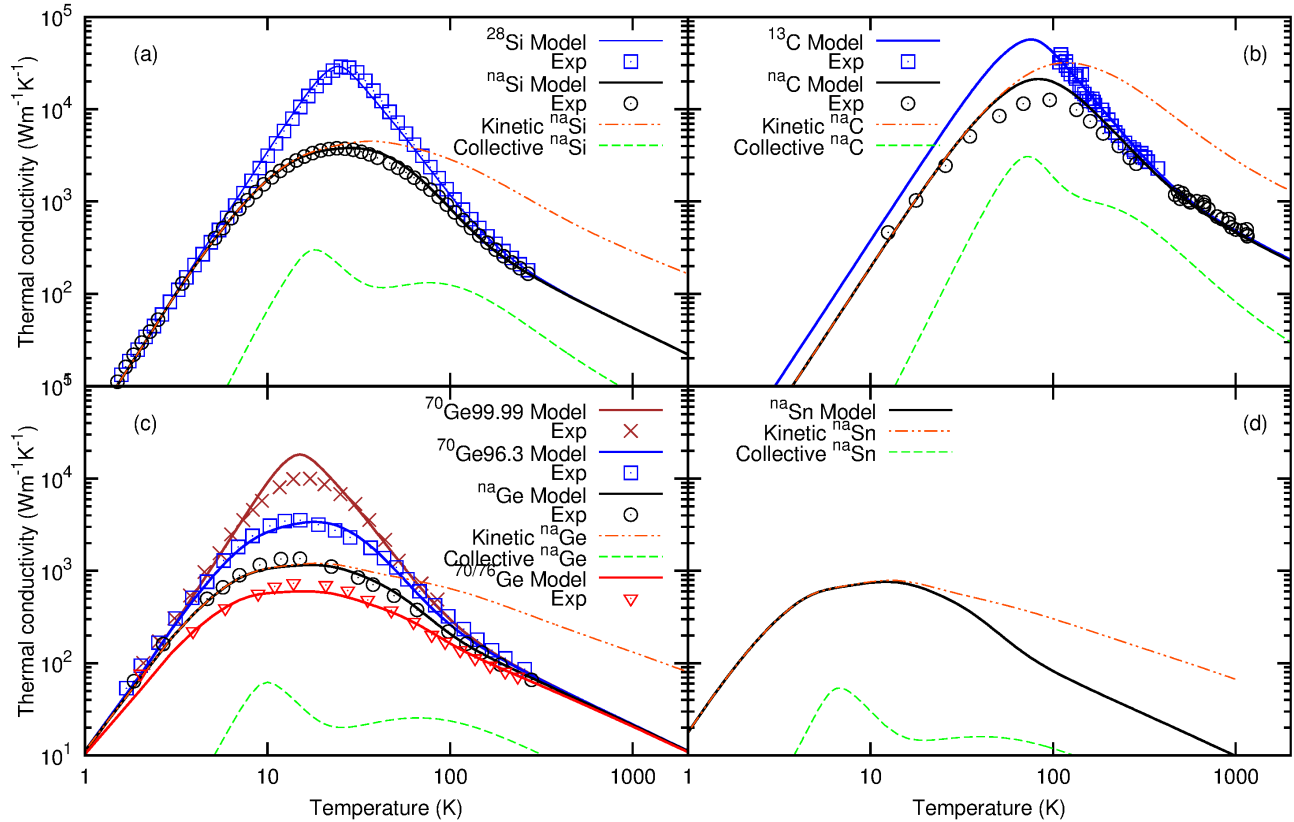}
    \end{center}
    \caption{(Color online) Thermal conductivity of the set of group-IV semiconductors. The fit for $^{\rm na}$Si (solid line) is shown in (a). Predictions (solid lines) are shown for $^{28}$Si in (a), for diamond samples in (b), Ge samples in (c) and $\alpha$-Sn in (d). The kinetic (dashed-dot lines) and collective (dashed lines) contributions are specified in each case. Experimental data\cite{Inyushkin2004,Asen-Palmer1997,Onn1992,Wei1993} appear in symbols. }\label{fig:FIT1}
\end{figure*}

In Figures \ref{fig:FIT1}(a)-(d) we show our theoretical predictions for Si, Ge, C and $\alpha$-Sn samples compared with experimental data\cite{Inyushkin2004,Asen-Palmer1997,Onn1992,Wei1993}. Note the remarkable agreement with measurements. These results are better than expected, as it is widely known that some of the materials properties (like $V$ or $v_g$) can have some dependence on $T$. This makes our phenomenological expressions expected to be valid only to obtain general trends in the behavior of $\kappa$.

In the plots it is also shown the kinetic and the collective limits for natural isotopic composition samples. It can be seen that as $T$ rises, $\kappa$ makes a transition from a purely kinetic to a collective behavior. This transition happens because at low temperature resistive boundary scattering is dominant and consequently the thermal transport is kinetic. As temperature rises, normal scattering is starting to be important and the collective behavior appears. The importance of this collective term is the same for all the samples. The only difference between them is the temperature this behavior is experienced at. The insufficient attention to this transition in usual theoretical models can be the cause for the difficulties in trying to obtain the same predictions by using the CM. Although CM and KCM give the same results in the limits ($\Sigma=1$ or $0$), the transition gives significantly different results.\cite{Krumhansl1965,Guyer1966} In the figure we have also included predictions for $\alpha$-Sn despite the lack of experimental values for this material. 

In this work we obviously do not pretend to obtain extremely good fits, as the expressions used can depend on specific characteristics of the dispersion relations for each material. In spite of this, from these results it can be deduced that usual relaxation times expressions are good enough to calculate thermal conductivity, and that some of the issues when fitting experimental values could be related to the phenomenological model. 

In conclusion, we have shown that with a kinetic-collective model it can be predicted the thermal conductivity of a group of samples of different materials without additional fitting parameters. The model allows to interpret the behavior of the thermal conductivity as a transition from a kinetic to a collective regime. The results can shed light to the understanding of the phonon-phonon interaction in this kind of samples as it can be interpreted in terms of these different behaviors.  

The authors acknowledge financial support from projects CSD2010-00044, FIS2012-32099, MAT2012-33483, and 2009-SGR00164, and from a Marie Curie Reintegration Grant. The authors thank Prof. D. Jou for a critical reading of the manuscript. Thanks are also given to the Red Espa\~{n}ola de Supercomputaci\'{o}n providing access to the supercomputer TIRANT.

\bibliographystyle{apsrev}

\begin{thebibliography}{26}
\expandafter\ifx\csname natexlab\endcsname\relax\def\natexlab#1{#1}\fi
\expandafter\ifx\csname bibnamefont\endcsname\relax
  \def\bibnamefont#1{#1}\fi
\expandafter\ifx\csname bibfnamefont\endcsname\relax
  \def\bibfnamefont#1{#1}\fi
\expandafter\ifx\csname citenamefont\endcsname\relax
  \def\citenamefont#1{#1}\fi
\expandafter\ifx\csname url\endcsname\relax
  \def\url#1{\texttt{#1}}\fi
\expandafter\ifx\csname urlprefix\endcsname\relax\def\urlprefix{URL }\fi
\providecommand{\bibinfo}[2]{#2}
\providecommand{\eprint}[2][]{\url{#2}}

\bibitem[{\citenamefont{Fugallo et~al.}(2013)\citenamefont{Fugallo, Lazzeri,
  Paulatto, and Mauri}}]{Fugallo2013}
\bibinfo{author}{\bibfnamefont{G.}~\bibnamefont{Fugallo}},
  \bibinfo{author}{\bibfnamefont{M.}~\bibnamefont{Lazzeri}},
  \bibinfo{author}{\bibfnamefont{L.}~\bibnamefont{Paulatto}}, \bibnamefont{and}
  \bibinfo{author}{\bibfnamefont{F.}~\bibnamefont{Mauri}},
  \bibinfo{journal}{Phys. Rev. B} \textbf{\bibinfo{volume}{88}},
  \bibinfo{pages}{045430} (\bibinfo{year}{2013}).

\bibitem[{\citenamefont{Allen}(2013)}]{Allen2013}
\bibinfo{author}{\bibfnamefont{P.~B.} \bibnamefont{Allen}},
  \bibinfo{journal}{Phys. Rev. B} \textbf{\bibinfo{volume}{88}},
  \bibinfo{pages}{144302} (\bibinfo{year}{2013}).

\bibitem[{\citenamefont{Callaway}(1959)}]{Callaway1959}
\bibinfo{author}{\bibfnamefont{J.}~\bibnamefont{Callaway}},
  \bibinfo{journal}{Phys. Rev.} \textbf{\bibinfo{volume}{113}},
  \bibinfo{pages}{1046} (\bibinfo{year}{1959}).

\bibitem[{\citenamefont{Guyer and Krumhansl}(1966)}]{Guyer1966}
\bibinfo{author}{\bibfnamefont{R.}~\bibnamefont{Guyer}} \bibnamefont{and}
  \bibinfo{author}{\bibfnamefont{J.}~\bibnamefont{Krumhansl}},
  \bibinfo{journal}{Phys. Rev.} \textbf{\bibinfo{volume}{148}},
  \bibinfo{pages}{778} (\bibinfo{year}{1966}).

\bibitem[{\citenamefont{Broido et~al.}(2007)\citenamefont{Broido, Malorny,
  Birner, Mingo, and Stewart}}]{Broido2007}
\bibinfo{author}{\bibfnamefont{D.~A.} \bibnamefont{Broido}},
  \bibinfo{author}{\bibfnamefont{M.}~\bibnamefont{Malorny}},
  \bibinfo{author}{\bibfnamefont{G.}~\bibnamefont{Birner}},
  \bibinfo{author}{\bibfnamefont{N.}~\bibnamefont{Mingo}}, \bibnamefont{and}
  \bibinfo{author}{\bibfnamefont{D.~A.} \bibnamefont{Stewart}},
  \bibinfo{journal}{Appl. Phys. Lett.} \textbf{\bibinfo{volume}{91}},
  \bibinfo{pages}{231922} (\bibinfo{year}{2007}).

\bibitem[{\citenamefont{Asen-Palmer et~al.}(1997)\citenamefont{Asen-Palmer,
  Bartkowski, Gmelin, Cardona, Zhernov, Inyushkin, Taldenkov, Ozhogin, Itoh,
  and Haller}}]{Asen-Palmer1997}
\bibinfo{author}{\bibfnamefont{M.}~\bibnamefont{Asen-Palmer}},
  \bibinfo{author}{\bibfnamefont{K.}~\bibnamefont{Bartkowski}},
  \bibinfo{author}{\bibfnamefont{E.}~\bibnamefont{Gmelin}},
  \bibinfo{author}{\bibfnamefont{M.}~\bibnamefont{Cardona}},
  \bibinfo{author}{\bibfnamefont{A.~P.} \bibnamefont{Zhernov}},
  \bibinfo{author}{\bibfnamefont{A.~V.} \bibnamefont{Inyushkin}},
  \bibinfo{author}{\bibfnamefont{A.}~\bibnamefont{Taldenkov}},
  \bibinfo{author}{\bibfnamefont{V.~I.} \bibnamefont{Ozhogin}},
  \bibinfo{author}{\bibfnamefont{K.~M.} \bibnamefont{Itoh}}, \bibnamefont{and}
  \bibinfo{author}{\bibfnamefont{E.~E.} \bibnamefont{Haller}},
  \bibinfo{journal}{Phys. Rev. B} \textbf{\bibinfo{volume}{56}},
  \bibinfo{pages}{9431} (\bibinfo{year}{1997}).

\bibitem[{\citenamefont{Esfarjani et~al.}(2011)\citenamefont{Esfarjani, Chen,
  and Stokes}}]{Esfarjani2011}
\bibinfo{author}{\bibfnamefont{K.}~\bibnamefont{Esfarjani}},
  \bibinfo{author}{\bibfnamefont{G.}~\bibnamefont{Chen}}, \bibnamefont{and}
  \bibinfo{author}{\bibfnamefont{H.~T.} \bibnamefont{Stokes}},
  \bibinfo{journal}{Phys. Rev. B} \textbf{\bibinfo{volume}{84}},
  \bibinfo{pages}{085204} (\bibinfo{year}{2011}).

\bibitem[{\citenamefont{Li and Mingo}(2013)}]{Li2013}
\bibinfo{author}{\bibfnamefont{W.}~\bibnamefont{Li}} \bibnamefont{and}
  \bibinfo{author}{\bibfnamefont{N.}~\bibnamefont{Mingo}}, \bibinfo{journal}{J.
  Appl. Phys.} \textbf{\bibinfo{volume}{114}}, \bibinfo{pages}{183505}
  (\bibinfo{year}{2013}).

\bibitem[{\citenamefont{Morelli et~al.}(2002)\citenamefont{Morelli, Heremans,
  and Slack}}]{Morelli2002}
\bibinfo{author}{\bibfnamefont{D.}~\bibnamefont{Morelli}},
  \bibinfo{author}{\bibfnamefont{J.}~\bibnamefont{Heremans}}, \bibnamefont{and}
  \bibinfo{author}{\bibfnamefont{G.}~\bibnamefont{Slack}},
  \bibinfo{journal}{Phys. Rev. B} \textbf{\bibinfo{volume}{66}},
  \bibinfo{pages}{195304} (\bibinfo{year}{2002}).

\bibitem[{\citenamefont{Mingo}(2003)}]{Mingo2003}
\bibinfo{author}{\bibfnamefont{N.}~\bibnamefont{Mingo}},
  \bibinfo{journal}{Phys. Rev. B} \textbf{\bibinfo{volume}{68}},
  \bibinfo{pages}{113308} (\bibinfo{year}{2003}).

\bibitem[{\citenamefont{Chantrenne et~al.}(2005)\citenamefont{Chantrenne,
  Barrat, Blase, and J.D.Gale}}]{Chantrenne2005}
\bibinfo{author}{\bibfnamefont{P.}~\bibnamefont{Chantrenne}},
  \bibinfo{author}{\bibfnamefont{J.~L.} \bibnamefont{Barrat}},
  \bibinfo{author}{\bibfnamefont{X.}~\bibnamefont{Blase}}, \bibnamefont{and}
  \bibinfo{author}{\bibnamefont{J.D.Gale}}, \bibinfo{journal}{J. App. Phys.}
  \textbf{\bibinfo{volume}{97}}, \bibinfo{pages}{104318}
  (\bibinfo{year}{2005}).

\bibitem[{\citenamefont{Alvarez et~al.}(2010)\citenamefont{Alvarez,
  Alvarez-Quintana, Jou, and Viejo}}]{Alvarez2010}
\bibinfo{author}{\bibfnamefont{F.~X.} \bibnamefont{Alvarez}},
  \bibinfo{author}{\bibfnamefont{J.}~\bibnamefont{Alvarez-Quintana}},
  \bibinfo{author}{\bibfnamefont{D.}~\bibnamefont{Jou}}, \bibnamefont{and}
  \bibinfo{author}{\bibfnamefont{J.~R.} \bibnamefont{Viejo}},
  \bibinfo{journal}{J. Appl. Phys.} \textbf{\bibinfo{volume}{107}},
  \bibinfo{pages}{084303} (\bibinfo{year}{2010}).

\bibitem[{\citenamefont{Ward et~al.}(2009)\citenamefont{Ward, Broido, Stewart,
  and Deinzer}}]{Ward2009}
\bibinfo{author}{\bibfnamefont{A.}~\bibnamefont{Ward}},
  \bibinfo{author}{\bibfnamefont{D.}~\bibnamefont{Broido}},
  \bibinfo{author}{\bibfnamefont{D.}~\bibnamefont{Stewart}}, \bibnamefont{and}
  \bibinfo{author}{\bibfnamefont{G.}~\bibnamefont{Deinzer}},
  \bibinfo{journal}{Phys. Rev. B} \textbf{\bibinfo{volume}{80}},
  \bibinfo{pages}{125203} (\bibinfo{year}{2009}).

\bibitem[{\citenamefont{Kazan et~al.}(2010)\citenamefont{Kazan, Guisbiers,
  Pereira, Correia, Masri, Bruyant, Volz, and Royer}}]{Kazan2010}
\bibinfo{author}{\bibfnamefont{M.}~\bibnamefont{Kazan}},
  \bibinfo{author}{\bibfnamefont{G.}~\bibnamefont{Guisbiers}},
  \bibinfo{author}{\bibfnamefont{S.}~\bibnamefont{Pereira}},
  \bibinfo{author}{\bibfnamefont{M.~R.} \bibnamefont{Correia}},
  \bibinfo{author}{\bibfnamefont{P.}~\bibnamefont{Masri}},
  \bibinfo{author}{\bibfnamefont{A.}~\bibnamefont{Bruyant}},
  \bibinfo{author}{\bibfnamefont{S.}~\bibnamefont{Volz}}, \bibnamefont{and}
  \bibinfo{author}{\bibfnamefont{P.}~\bibnamefont{Royer}}, \bibinfo{journal}{J.
  Appl. Phys.} \textbf{\bibinfo{volume}{107}}, \bibinfo{pages}{83503}
  (\bibinfo{year}{2010}).

\bibitem[{\citenamefont{de~Tomas et~al.}(2013)\citenamefont{de~Tomas,
  Cantarero, Lopeandia, and Alvarez}}]{DeTomas2013}
\bibinfo{author}{\bibfnamefont{C.}~\bibnamefont{de~Tomas}},
  \bibinfo{author}{\bibfnamefont{A.}~\bibnamefont{Cantarero}},
  \bibinfo{author}{\bibfnamefont{A.~F.} \bibnamefont{Lopeandia}},
  \bibnamefont{and} \bibinfo{author}{\bibfnamefont{F.~X.}
  \bibnamefont{Alvarez}} (\bibinfo{year}{2013}), \eprint{arXiv:1310.7127
  [cond-mat.mes-hall]}.

\bibitem[{\citenamefont{Weber}(1974)}]{Weber1974}
\bibinfo{author}{\bibfnamefont{W.}~\bibnamefont{Weber}},
  \bibinfo{journal}{Phys. Rev. Lett.} \textbf{\bibinfo{volume}{33}},
  \bibinfo{pages}{371} (\bibinfo{year}{1974}).

\bibitem[{\citenamefont{Casimir}(1938)}]{Casimir1938}
\bibinfo{author}{\bibfnamefont{H.~B.~G.} \bibnamefont{Casimir}},
  \bibinfo{journal}{Physica} \textbf{\bibinfo{volume}{5}}, \bibinfo{pages}{595}
  (\bibinfo{year}{1938}).

\bibitem[{\citenamefont{Klemens}(1955)}]{Klemens1955}
\bibinfo{author}{\bibfnamefont{P.~G.} \bibnamefont{Klemens}},
  \bibinfo{journal}{Proc. Phys. Soc. London, Sect. A}
  \textbf{\bibinfo{volume}{68}}, \bibinfo{pages}{1113} (\bibinfo{year}{1955}).

\bibitem[{\citenamefont{Tamura}(1983)}]{Tamura1983}
\bibinfo{author}{\bibfnamefont{S.}~\bibnamefont{Tamura}},
  \bibinfo{journal}{Phys. Rev. B} \textbf{\bibinfo{volume}{27}},
  \bibinfo{pages}{858} (\bibinfo{year}{1983}).

\bibitem[{\citenamefont{Inyushkin et~al.}(2004)\citenamefont{Inyushkin,
  Taldenkov, Gibin, Gusev, and Pohl}}]{Inyushkin2004}
\bibinfo{author}{\bibfnamefont{A.~V.} \bibnamefont{Inyushkin}},
  \bibinfo{author}{\bibfnamefont{A.~N.} \bibnamefont{Taldenkov}},
  \bibinfo{author}{\bibfnamefont{A.~M.} \bibnamefont{Gibin}},
  \bibinfo{author}{\bibfnamefont{A.~V.} \bibnamefont{Gusev}}, \bibnamefont{and}
  \bibinfo{author}{\bibfnamefont{H.-J.} \bibnamefont{Pohl}},
  \bibinfo{journal}{Phys. Status Solidi C} \textbf{\bibinfo{volume}{1}},
  \bibinfo{pages}{2995} (\bibinfo{year}{2004}).

\bibitem[{\citenamefont{Wei et~al.}(1993)\citenamefont{Wei, Kuo, Thomas,
  Anthony, and Banholzer}}]{Wei1993}
\bibinfo{author}{\bibfnamefont{L.}~\bibnamefont{Wei}},
  \bibinfo{author}{\bibfnamefont{P.}~\bibnamefont{Kuo}},
  \bibinfo{author}{\bibfnamefont{R.}~\bibnamefont{Thomas}},
  \bibinfo{author}{\bibfnamefont{T.}~\bibnamefont{Anthony}}, \bibnamefont{and}
  \bibinfo{author}{\bibfnamefont{W.}~\bibnamefont{Banholzer}},
  \bibinfo{journal}{Phys. Rev. Lett.} \textbf{\bibinfo{volume}{70}},
  \bibinfo{pages}{3764} (\bibinfo{year}{1993}).

\bibitem[{\citenamefont{Onn et~al.}(1992)\citenamefont{Onn, Witek, Qiu,
  Anthony, and Banholzer}}]{Onn1992}
\bibinfo{author}{\bibfnamefont{D.}~\bibnamefont{Onn}},
  \bibinfo{author}{\bibfnamefont{A.}~\bibnamefont{Witek}},
  \bibinfo{author}{\bibfnamefont{Y.}~\bibnamefont{Qiu}},
  \bibinfo{author}{\bibfnamefont{T.}~\bibnamefont{Anthony}}, \bibnamefont{and}
  \bibinfo{author}{\bibfnamefont{W.}~\bibnamefont{Banholzer}},
  \bibinfo{journal}{Phys. Rev. Lett.} \textbf{\bibinfo{volume}{68}},
  \bibinfo{pages}{2806} (\bibinfo{year}{1992}).

\bibitem[{\citenamefont{Ward and Broido}(2010)}]{Ward2010}
\bibinfo{author}{\bibfnamefont{A.}~\bibnamefont{Ward}} \bibnamefont{and}
  \bibinfo{author}{\bibfnamefont{D.~A.} \bibnamefont{Broido}},
  \bibinfo{journal}{Phys. Rev. B} \textbf{\bibinfo{volume}{81}},
  \bibinfo{pages}{085205} (\bibinfo{year}{2010}).

\bibitem[{\citenamefont{Leibfried and Schl\"{o}mann}(1954)}]{Leibfried1954}
\bibinfo{author}{\bibfnamefont{G.}~\bibnamefont{Leibfried}} \bibnamefont{and}
  \bibinfo{author}{\bibfnamefont{E.}~\bibnamefont{Schl\"{o}mann}},
  \bibinfo{journal}{Nachr. Akad. Wiss. G\"{o}ttingen}
  \textbf{\bibinfo{volume}{IIa(4)}}, \bibinfo{pages}{71}
  (\bibinfo{year}{1954}).

\bibitem[{\citenamefont{Ehrenreich et~al.}(1979)\citenamefont{Ehrenreich,
  Seitz, Turnbull, and Slack}}]{Slack1973}
\bibinfo{author}{\bibfnamefont{H.}~\bibnamefont{Ehrenreich}},
  \bibinfo{author}{\bibfnamefont{F.}~\bibnamefont{Seitz}},
  \bibinfo{author}{\bibfnamefont{D.}~\bibnamefont{Turnbull}}, \bibnamefont{and}
  \bibinfo{author}{\bibfnamefont{G.~A.} \bibnamefont{Slack}},
  \bibinfo{journal}{Solid State Physics} \textbf{\bibinfo{volume}{34}},
  \bibinfo{pages}{1} (\bibinfo{year}{1979}).

\bibitem[{\citenamefont{Krumhansl}(1965)}]{Krumhansl1965}
\bibinfo{author}{\bibfnamefont{J.~A.} \bibnamefont{Krumhansl}},
  \bibinfo{journal}{Proc. Phys. Soc.} \textbf{\bibinfo{volume}{85}},
  \bibinfo{pages}{921} (\bibinfo{year}{1965}).

\end{thebibliography}

\end{document}